\documentclass[prc,aps,floats,floatfix,twocolumn,showpacs,nofootinbib,%
superscriptaddress]{revtex4}

\usepackage{amsmath}
\usepackage{amssymb}
\usepackage{amsfonts}
\usepackage{graphicx}
\usepackage{tabularx}
\usepackage{multirow}
\usepackage{lscape}
\usepackage{rotating}

\usepackage{pstricks}

\renewcommand{\vec}[1]{{\mathbf #1}}
\newcommand{\etal}{\emph{et~al.}}
\newcommand{\nn}{\nonumber}
\newcommand{\mcc}[1]{\multicolumn{1}{c}{#1}}

\newcommand{\vnabla}{\boldsymbol{\nabla}}
\newcommand{\vsigma}{\boldsymbol{\sigma}}

\newcommand{\wt}{\widetilde}
\newcommand{\eps}{\varepsilon}
\newcommand{\alp}{\alpha}
\newcommand{\cogam}{(\gamma + 1) (\gamma + 2)}

\newcommand{\rhozg}{\rho_0^{\gamma}}

\newcommand{\wzzu}{W^{(0,0)}_1}
\newcommand{\wuzu}{W^{(1,0)}_1}
\newcommand{\wzuu}{W^{(0,1)}_1}
\newcommand{\wuuu}{W^{(1,1)}_1}
\newcommand{\wzzd}{W^{(0,0)}_2}
\newcommand{\wuzd}{W^{(1,0)}_2}
\newcommand{\wzud}{W^{(0,1)}_2}
\newcommand{\wuud}{W^{(1,1)}_2}
\newcommand{\wsod}{W_0^2}
\newcommand{\wua}{W^{(\alp)}_1}
\newcommand{\wda}{W^{(\alp)}_2}
\newcommand{\wtua}{{\wt W}^{(\alp)}_1}
\newcommand{\bdbt}{\beta_2-\beta_3}

\newcommand{\chihf}{\chi_{HF}}
\newcommand{\chiz}{\chi_0}
\newcommand{\chid}{\chi_2}
\newcommand{\chiq}{\chi_4}
\newcommand{\chia}{\chi^{(\alp)}}
\newcommand{\chizd}{\chiz^2}
\newcommand{\chidd}{\chid^2}
\newcommand{\cchi}{\chiz \chiq}
\newcommand{\tm}{(t_e-t_o)}
\newcommand{\tp}{(t_e+t_o)}
\newcommand{\tmc}{(t_e-5t_o)}
\newcommand{\tpc}{(t_e+5t_o)}
\newcommand{\kud}{k_{12}}
\newcommand{\kuz}{(\kud)^{(1)}_0}
\newcommand{\kum}{(\kud)^{(1)}_{\text{-}1}}
\newcommand{\kuu}{(\kud)^{(1)}_1}
\newcommand{\brk}{\left[ \kuz \kuz + \kum \kuu \right]}

\newcommand{\pquar}{\tfrac{1}{4}}
\newcommand{\demi}{\frac{1}{2}}
\newcommand{\pdemi}{\tfrac{1}{2}}
\newcommand{\tdemi}{\frac{3}{2}}
\newcommand{\ptdemi}{\tfrac{3}{2}}
\newcommand{\tqpi}{\frac{3}{4\pi}}
\newcommand{\thpi}{\frac{3}{8\pi}}

\newcommand{\moq}{\frac{m^*\omega}{q}}
\newcommand{\mokf}{\frac{m^*\omega}{k_F^2}}
\newcommand{\moqd}{\left( \moq \right)^2}
\newcommand{\mokfd}{\left( \mokf \right)^2}

\newcommand{\mkdpi}{\frac{m^* k_F}{2\pi^2}}
\newcommand{\mktpi}{\frac{m^* k_F}{3\pi^2}}
\newcommand{\mkspi}{\frac{m^* k_F}{6\pi^2}}
\newcommand{\mkcpi}{\frac{m^* k_F^3}{\pi^2}}
\newcommand{\mkcdpi}{\frac{m^* k_F^3}{2\pi^2}}
\newcommand{\mkctpi}{\frac{m^* k_F^3}{3\pi^2}}

\newcommand{\mkfpi}{\frac{m^* k_F^5}{\pi^2}}
\newcommand{\mkfdpi}{\frac{m^* k_F^5}{2\pi^2}}
\newcommand{\dmkctpi}{\frac{2m^* k_F^3}{3\pi^2}}
\newcommand{\mspikq}{\frac{m^*}{6\pi ^2k_F} q^2} 
\newcommand{\crochi}{\left[\chidd-\cchi+\mokfd \chizd-\mspikq \chiz \right]}
\newcommand{\Xa}{X^{(1, 0,\pm 1)}}
\newcommand{\ttpXa}{\wuzd + 2 \tmc + \Xa}
\newcommand{\ttmXa}{\wuzd + 2 \tmc - \Xa}
\newcommand{\Xb}{X^{(1, 0, 0)}}
\newcommand{\ttpXb}{\wuzd - \tmc + \Xb}
\newcommand{\ttmXb}{\wuzd - \tmc - \Xb}
\newcommand{\Xc}{X^{(1, 1,\pm 1)}}
\newcommand{\ttpXc}{\wuud + 6 \tm + \Xc}
\newcommand{\ttmXc}{\wuud + 6 \tm - \Xc}
\newcommand{\Xd}{X^{(1, 1, 0)}}
\newcommand{\ttpXd}{\wuud - 3 \tm + \Xd}
\newcommand{\ttmXd}{\wuud - 3 \tm - \Xd}

\newcommand{\romq}{\mathrm q}

\newcommand{\br}{\vec{r}}
\newcommand{\bR}{\vec{R}}
\newcommand{\bk}{{\bf k}}
\newcommand{\bku}{\bk_1}
\newcommand{\bkd}{\bk_2}
\newcommand{\bq}{{\bf q}}
\newcommand{\bqt}{\bq_3}
\newcommand{\ghf}{G_{HF}(q,\omega,\bku)}
\newcommand{\grpau}{G_{RPA}^{(\alp)}(q,\omega,\bku)}
\newcommand{\grpad}{G_{RPA}^{(\alp')}(q,\omega,\bkd)}
\newcommand{\vph}{V_{ph}^{(\alp,\alp')}(q,\bku,\bkd)}
\newcommand{\dd}{\delta (\br)}

\newcommand{\la}{\langle \,}
\newcommand{\ra}{\, \rangle}

\newcommand{\be}{\begin{equation}}
\newcommand{\ee}{\end{equation}}
\newcommand{\bqr}{\begin{eqnarray}}
\newcommand{\eqr}{\end{eqnarray}}
\newcommand{\bi}{\begin{itemize}}
\newcommand{\ei}{\end{itemize}}
\newcommand{\bwt}{\begin{widetext}}
\newcommand{\ewt}{\end{widetext}}

\newcommand{\citeqdot}[1]{eq.~(\ref{#1})}

\newcommand{\citeqssdot}[3]{eqs.~(\ref{#1}), (\ref{#2}) and~(\ref{#3})}
\newcommand{\citerefdot}[1]{ref.\cite{#1}}
\newcommand{\citerefsdot}[1]{refs.\cite{#1}}

\newcommand{\citeFigure}[1]{Figure~\ref{#1}}

\newcommand{\citeAppendix}[1]{Appendix~\ref{#1}}

\newcommand{\citeTable}[1]{Table~\ref{#1}}

\begin{document}

\title{Nuclear response for the Skyrme effective interaction with
       zero-range tensor terms}


\author{D. Davesne}
\email{davesne@ipnl.in2p3.fr}
\affiliation{Universit\'e de Lyon, F-69003 Lyon, France; Universit\'e Lyon 1,
             43 Bd. du 11 Novembre 1918, F-69622 Villeurbanne cedex, France\\
             CNRS-IN2P3, UMR 5822, Institut de Physique Nucl{\'e}aire de Lyon}

\author{M. Martini}
\email{martini@ipnl.in2p3.fr}
\affiliation{Universit\'e de Lyon, F-69003 Lyon, France; Universit\'e Lyon 1,
             43 Bd. du 11 Novembre 1918, F-69622 Villeurbanne cedex, France\\
             CNRS-IN2P3, UMR 5822, Institut de Physique Nucl{\'e}aire de Lyon}
\affiliation{Universit\`a di Bari, I-70126 Bari, Italia}

\author{K. Bennaceur}
\email{bennaceur@ipnl.in2p3.fr}
\affiliation{Universit\'e de Lyon, F-69003 Lyon, France; Universit\'e Lyon 1,
             43 Bd. du 11 Novembre 1918, F-69622 Villeurbanne cedex, France\\
             CNRS-IN2P3, UMR 5822, Institut de Physique Nucl{\'e}aire de Lyon}

\author{J. Meyer}
\email{jmeyer@ipnl.in2p3.fr}
\affiliation{Universit\'e de Lyon, F-69003 Lyon, France; Universit\'e Lyon 1,
             43 Bd. du 11 Novembre 1918, F-69622 Villeurbanne cedex, France\\
             CNRS-IN2P3, UMR 5822, Institut de Physique Nucl{\'e}aire de Lyon}


\begin{abstract}
The effects of a zero-range tensor component of the effective
interaction on nuclear response functions are determined in the so-called
RPA approach.
Explicit formula in the case of symmetric homogeneous isotropic
nuclear matter are given for each spin-isospin excitation channel. 
It is shown for a typical interaction
with tensor couplings that the effects are quantitatively important,
mainly in vector channels.
\end{abstract}


 
\date{\today}


\maketitle


\section{Introduction}
\label{sect:intro}


Microscopic mean-field approaches are the only ones that allow 
for systematic calculations of binding energy and one 
body observables in the region of the nuclear chart that 
ranges from medium to heavy mass atomic nuclei from drip 
line to drip line~\cite{RMP}. 
These effective approaches rely on a limited number of 
universal parameters, usually fitted on experimental data
along with properties of infinite nuclear matter derived
from realistic models~\cite{meyer03}.

In the Skyrme-Hartree-Fock formulation, the energy of 
the system takes the form of a functional of local one 
body densities derived from the effective Skyrme 
interaction {\em ansatz}. 
The commonly used Skyrme interaction typically depends 
on ten parameters and is made of contact central terms 
including a spin-orbit interaction.
The latter, which is controlled by one single parameter, 
is mandatory to obtain the known sequence of magic numbers 
along the valley of stability.
Although this is enough to reproduce the global features 
of nuclei, it was stressed from the beginning by Skyrme 
himself~\cite{Sky56a,Bel56a} that such a simple interaction 
would probably not be sufficient for a realistic 
description of nuclear spectroscopy
and that a tensor interaction might be needed~\cite{Sky59a}.
This last part of the effective interaction, made of two
contact terms, was
not considered in most of the early
parametrizations of the Skyrme force
possibly because of the difficulty to constrain the 
corresponding coupling constants.

In spite of the difficulties related to the adjustment
of the parameters of the tensor terms, over the years, several attemps have
been made for including them.
The tensor terms as proposed by Skyrme
were considered on top of the Skyrme SIII~\cite{Bei75a} 
effective interaction~\cite{Sta77a} or with a complete refit
of the parameters~\cite{Ton83a,Liu91a}.
More recently, the tensor effective interaction has regained some
attention~\cite{Bro06a,Col07a,Bri07a,Zalewski08,Bai09a,Bai09b}, partly 
because it was
supposed to be the key for the reproduction of several specific
spectroscopic features like, for example, the relative shift
of proton $1g_{7/2}$ and $1h_{11/2}$ levels
in antimony isotopes~\cite{Schi04aE}.
In a recent article~\cite{Lesinski07}, a systematic study of the zero
range effective tensor interaction combined with
a standard Skyrme functional has been made. In this work,
a set of interactions was built by fixing the two parameters of
the tensor terms to different values while the remaining part of the
interactions was fitted
using the same procedure as for the well known SLy
interaction~\cite{Cha97a,Cha98a,Cha98b}.
It was shown that global features of spherical nuclei (masses and
radii) and single-particle energies cannot be used to clearly mark
the boundary of a successful domain for the tensor parameters.

In addition of looking for the parameters that lead to the ``best fits'' 
to the data, one should also worry about the values that could lead to
unphysical instabilities of nuclei.
The tensor interaction energy depending on spins and gradients of densities
of the interacting nucleons, one can intuitively understand that the
appearance of unphysical finite size domains of polarized nuclear matter
can be favored for some values of the coupling constants.
Such situations are hard to predict, and to avoid, during the fit 
of the parameters since polarized systems break time reversal
symmetry while the calculations entering the standard fitting
protocol assume that nuclei are spherical and time even.

A similar kind of instabilities
of finite systems was encountered and examined in an article devoted
to the study of effective mass splitting~\cite{Les06a}.
It was shown that the linear response formalism applied to the
Skyrme energy functional can be used to predict the appearance
of finite-size instabilities in nuclei.
However, only the central part of the Skyrme interaction was
taken into account for the building of the linear response.
In the present article, we derive the full linear response from the
energy calculated from a Skyrme effective interaction that
contains spin-orbit and tensor terms. More specifically,
we always consider that the energy is derived
from an effective interaction contrary to the spirit
of the Skyrme energy density functional method (Skyrme EDF)
for which this link is not required.

A significant effort is nowadays devoted to the introduction of new 
terms in the Skyrme EDF, the tensor term being
one~\cite{Bro06a,Lesinski07,Zalewski08},
other popular choices being new density dependent 
couplings~\cite{Krewald77,Ton84a,Farine01,Cochet04a,Cochet04b} or higher 
order derivative terms~\cite{Carlsson08}.
The number of parameters becoming larger, it is of particular importance 
to define a clever fitting strategy.
Acceptable ranges of variation for the parameters can be motivated by linking
the Skyrme EDF to realistic interactions using many-body perturbation
theory on top of renormalized low momentum interactions~\cite{Bog03,Duguet03}.
It is also obviously mandatory to locate the regions in parameters space
that lead to instabilities.
This is particularly crucial for the spin channels, since the corresponding
instabilities can only develop when the parameters are
plugged in calculation codes that allow for the
breaking of time reversal symmetry.

The linear response function is the tool of choice
which allows us to avoid areas of instabilities. This application will
be presented in a forthcoming article, the present one being mainly devoted
to the derivation of the general formulae.

This work is organized as follows: section~\ref{sect:skyrme} summarizes the
components of the Skyrme interaction which includes a zero range tensor,
section~\ref{sect:response} recalls the standard formalism of the linear
response in nuclear matter and discusses its generalization to these new
tensor terms. In section~\ref{sect:results}, we present some numerical
calculations of responses. Finally, we discuss further possible developments
in the conclusion.


\section{Skyrme interaction with tensor terms}
\label{sect:skyrme}


%
The usual \emph{ansatz} for the Skyrme effective
interaction~\cite{Cha97a,Cha98a} leads to an energy density functional which
can be written as the sum of a kinetic term, the Skyrme potential energy 
functional that models the effective strong interaction in the particle-hole 
channel, a pairing energy functional, the Coulomb energy functional and 
correction terms to approximately remove the contribution from the
center of mass motion.
The functional discussed in this article being applied to
infinite nuclear matter without pairing, we only consider
the kinetic and Skyrme potential energy terms.

Throughout this work, we will use an effective Skyrme energy functional,
as written in~\citeqdot{eq:EF:full}, that
corresponds to an antisymmetrized density-dependent 
two-body vertex in the particle-hole channel of the strong interaction.
It can be decomposed into a central, spin-orbit and tensor contributions
\begin{equation}
v_{\mathrm{Skyrme}} = v_\mathrm{C} + v_\mathrm{LS} + v_\mathrm{T} \,.
\end{equation}


We will use the standard density-dependent central Skyrme force
\bqr
\label{eq:Skyrme:central}
v_\mathrm{C} (\bR,\br)
& = &  t_0 \, ( 1 + x_0 \hat{P}_\sigma ) \; \dd                    \\
& + & \tfrac{1}{6}\,t_{3}\,(1+x_{3}\hat{P}_\sigma)
                       \, \rho^{\gamma} (\bR) \; \dd  \nn \\
& + & \tfrac{1}{2} \, t_1 \, ( 1 + x_1 \hat{P}_\sigma )
      \big[ \bk^{\prime 2} \; \dd + \dd \; \bk^2 \big]              \nn \\
& + & t_2 \ ( 1 + x_2 \hat{P}_\sigma ) \, \bk^{\prime} \cdot \dd \; \bk\,, \nn
\eqr
with the usual shorthand notations for $\br$, $\bR$, $\bk$, $\bk'$
and $\hat{P}_\sigma$~\cite{RMP}.

We will also use the  most standard form of the spin-orbit interaction
\bqr
\label{eq:Skyrme:LS}
v_\mathrm{LS} (\br)
& = & i \, W_0 \, ( \vsigma_1 + \vsigma_2 ) \cdot\left[
  \bk^{\prime} \times \dd \; \bk\right]
\eqr
which is a special case of the one proposed by Bell and
Skyrme~\cite{Bel56a,Sky59b}. Finally,
the tensor part of the interaction is the one proposed by
Skyrme~\cite{Sky56a,Sky59a}
\bwt
\bqr
\label{eq:Skyrme:tensor}
v_\mathrm{T} (\vec{r})
& = & \pdemi \, t_e \;
    \Big\{
    \big[ 3 \,( \vsigma_1 \cdot \vec{k}' ) \, ( \vsigma_2 \cdot \vec{k}' )
          - ( \vsigma_1 \cdot \vsigma_2 ) \, \vec{k}^{\prime 2} \,
    \big] \; \delta (\vec{r})
   +  \delta (\vec{r}) \;
      \big[ 3 \, ( \vsigma_1 \cdot \vec{k} ) \, ( \vsigma_2 \cdot \vec{k} )
            -  ( \vsigma_1 \cdot \vsigma_2) \, \vec{k}^{2}
      \Big]
    \Big\}
   \nn \\
&  & + \,t_o \,
     \Big[
       3 \, ( \vsigma_1 \cdot \vec{k}' ) \, \delta (\vec{r}) \,
            ( \vsigma_2 \cdot \vec{k} )
      - ( \vsigma_1 \cdot \vsigma_2 ) \, \vec{k}' \cdot \,
        \delta (\vec{r}) \, \vec{k}
     \Big]\,,
\eqr
\ewt
for which the derivation of the contribution to the total energy functional
is discussed in detail in~\citerefsdot{Flothesis,Per04a}
as well as in~\citerefdot{Lesinski07} where the impact of such a tensor
interaction on the properties of spherical nuclei is investigated.


\section{Linear response formalism}
\label{sect:response}


The linear response function in nuclear matter has already been widely
developed mainly in the framework of Random Phase Approximation (RPA) based
on the use of an effective interaction~\cite{Garcia92}.
We adopt the presentation of the work of Margueron \etal~\cite{Margueron06}
which was devoted to the study of the contribution from the spin-orbit term
to the linear response.

We consider here the case of infinite matter as a nuclear medium at zero
temperature and unpolarized both in spin and isospin spaces. 
At the mean field level this system is described as an ensemble of independent 
nucleons moving in a self-consistent mean field generated from an
effective interaction treated in the Hartree-Fock (HF) approximation.
For a given density,
the momentum dependent HF mean field, or self-energy, determines the 
single-particle spectrum $\eps(k)$ and the Fermi level $\eps(k_F)$.

To calculate the response of the medium to an external field, it is
convenient to introduce the Green function, or p-h (particle-hole)
propagator $G^{(\alp)}(\bq,\omega,\bku)$.
As it is illustrated on~\citeFigure{fig:qk1k2}, $\bku$ and $\bkd$  
are the initial and final hole momenta respectively and $\bq$ is the 
transferred momentum. 
We denote by $\alp = (S,M;I,Q)$ the spin and isospin particle-hole
channels with $S=0$ ($S=1$) for the non spin-flip (spin-flip) channel, 
$I=0$ ($I=1$) the isoscalar (isovector) channel, $M$ and $Q$ being the 
quantum numbers related with the projection of the operators
$\hat S$ and $\hat I$ on the quantification axis. The latter is chosen,
as usual, as the $z$ axis along the direction of $\bq$.

\begin{figure}[htbp]
      \includegraphics[width=0.95\columnwidth,clip]{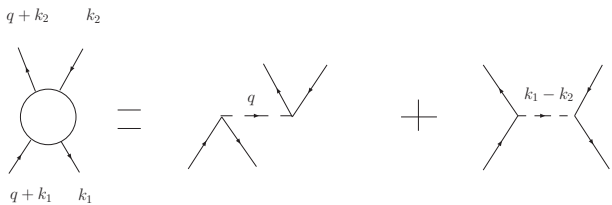}
      \caption{Direct and exchange parts of the p-h interaction.}
      \label{fig:qk1k2}
\end{figure}

In the HF approximation, the p-h Green function does not depend on 
the spin-isospin channel $(\alp)$ and reads~\cite{FW71}
\be 
   \ghf = \frac{\theta(k_F-k_1)-\theta(k_F - \vert \bku + \bq \vert)} 
        {\omega + \eps(k_1) - \eps(\vert \bku + \bq \vert) + i \eta \omega}\,.
\label{eq:GHF} 
\ee

To go beyond the HF approximation one takes into account 
long-range correlations by resumming a class of p-h
diagrams to obtain the well-known Random Phase Approximation~\cite{FW71}. 
The interaction appearing in the RPA is the p-h residual interaction 
whose matrix element including the exchange part can be written as
\be 
  \vph \equiv 
  \la \bq + \bku, \bku^{\text{-}1},(\alp) \vert V
   \vert \bq + \bkd, \bkd^{\text{-}1}, (\alp') \ra\,.
\label{eq:Vph}
\ee
In the general case, the residual interaction is obtained by taking
the second derivative of the total energy with respect to the densities
built from the Hartree-Fock solutions. In the absence of density
dependent terms, it can also be obtained by standard techniques
of particle-hole configuration~\cite{brussaard}.

The first important step is thus to determine the matrix elements
for the different parts of the p-h interaction
from~\citeqssdot{eq:Skyrme:central}{eq:Skyrme:LS}{eq:Skyrme:tensor}.


\subsection{The particle-hole interaction}


\subsubsection{Central part of the force}

The central component of the p-h interaction can be written in the
general form
\bwt
\be
   V_{\mathrm{C},ph}^{(\alp,\alp')}(q,\bku,\bkd)
    = \delta_{\alp\alp'}
   \left\{ \wua + \wda \left[ k_1^2 + k_2^2 \right]                   
                - \wda \frac{8 \pi}{3} k_1 k_2 
   \sum_{\mu=0,\pm 1} Y^{(1)\,*}_{\mu}(\hat{k}_1) 
              Y^{(1)}_{\mu}   (\hat{k}_2) 
   \right\} \, ,               
\label{eq:vph-c} 
\ee
\ewt
where the $W^{(\alp)}_1$ and $W^{(\alp)}_2$ coefficients are functions
of the Skyrme parameters $(t_i,x_i)$ and of the transferred momentum $\bq$
represented on figure~\ref{fig:qk1k2}. Their detailed expressions have
been given by Garcia-Recio~\etal~\cite{Garcia92}
and Navarro~\etal~\cite{Navarro99} for the symmetric nuclear matter
and pure neutron matter respectively while Hernandez~\etal~\cite{Hernandez97}
gave them for an arbitrary neutron-proton asymmetry.
The case of symmetric nuclear matter studied here is recalled in
the~\citeAppendix{app:sec:wcoef}.
The central part of the Skyrme interaction being usually density
dependent, it is not trivialy related to the p-h interaction
since the $W^{(\alp)}_i$ coefficients contain rearrangement terms.
One can note at this level, using only the central part of the interaction,
that there is no coupling between the different spin and isospin channels.


\subsubsection{Spin-orbit part of the force}

To calculate the contribution of the spin-orbit term
(see~\citeqdot{eq:Skyrme:LS}) to the p-h interaction one has to evaluate 
the matrix element of the spin-orbit interaction. 
Since this term is density-independent there is no rearrangement contribution 
and the result is just adding the following term to~\citeqdot{eq:vph-c}
(see Margueron \etal~\cite{Margueron06}):
\begin{align}
V_{\mathrm{LS},ph}^{(\alp,\alp')}(q,&\bku,\bkd)=
   - \delta_{II'} \, w(I) \, \sqrt{\frac{4 \pi}{3}} \, q \, W_0\,  \\
  \times\bigg\{ &   \delta_{S1} \delta_{S'0} M
        \left[ k_1 Y^{(1)}_{\text{-}M} (\hat{k}_1)
             - k_2 Y^{(1)}_{\text{-}M} (\hat{k}_2) \right]             \nn \\
  +&\,  \delta_{S0} \delta_{S'1} M^\prime
        \left[ k_1 Y^{(1)}_{M^\prime} (\hat{k}_1)
             - k_2 Y^{(1)}_{M^\prime} (\hat{k}_2)  \right]
        \bigg\}~,                                                      \nn
\label{eq:vph-so}
\end{align}
the factor $w(I)=3$ for $I=0$ and $w(I)=1$ for $I=1$ in the
case of symmetric nuclear matter. 
It is clear from this expression that the main effect of the 
spin-orbit component is to couple the $S=0$ and $S=1$ channels.


\subsubsection{Tensor part of the force}

With the tensor force previously defined (see~\citeqdot{eq:Skyrme:tensor}),
we have to calculate the antisymmetrized particle-hole matrix elements 
$\la IQ SM ; ph\, \vert v_\mathrm{T} \vert\, I'Q' S'M' ; ph \ra$.
Their analytical expressions
are summarized in \citeTable{tab:phme_tens} where we have adopted the 
following notation:
\be
   (k_{12})^{(1)}_M \equiv \sqrt{\frac{4\pi}{3}} 
                       \left[ k_1 Y^{(1)}_{M} (\hat{k}_1) 
                            - k_2 Y^{(1)}_{M} (\hat{k}_2) 
                       \right]\,.
\ee

Even if one can note from \citeTable{tab:phme_tens} that
channels with different spin projection $M$
are now coupled in a non trivial way, these additional matrix elements
are still diagonal in isospin space and act only in the vector channel.
However, since we include both spin-orbit and tensor interactions
in our approach,  it is fundamental to note that the tensor component
will impact both scalar and vector channels {\it via} the spin-orbit term.


\subsection{Response function}

With the particle-hole matrix elements we are now in position to solve 
the RPA problem itself, that is the Bethe-Salpeter equation satisfied by 
the RPA correlated Green function $\grpau$:
\bqr
   \grpau & = & \ghf + \ghf \, \nn \\
          &   & 
   \times \sum_{(\alp')} \, \int \frac{{\mathrm d}^3k_2}{(2 \pi)^3}
   \, \vph                     \nn \\
          &   &  
   \times \, \grpad\,.
\label{eq:eqBS} 
\eqr
The response function $\chi^{(\alp)}(q,\omega)$ in the infinite medium 
is related to the p-h Green function by:
\be 
   \chi_{RPA}^{(\alp)}(q,\omega)
       \, = \, g \int\frac{{\mathrm d}^3 k_1}{(2 \pi)^3} \, \grpau~,
\label{eq:chi} 
\ee
where the spin-isospin degeneracy factor $g$ is 4 for symmetric nuclear matter.
The Lindhard function $\chi_{HF}$ is obtained when the free p-h
propagator $G_{HF}$ is used in~\citeqdot{eq:chi}.

Following the notation of~\citerefsdot{Garcia92,Margueron06}, we define 
for any function $f(q,\omega,k_1)$:
\be
 \la f(\mathbf k_1)\ra
     \,\equiv\,\int \frac{\mathrm d^3 \bk_1}{(2\pi)^3} \, f(k_1)\,.
\ee
The response function can thus be written in each channel $(\alp)$ as:
\be
  \chi_{RPA}^{(\alp)} \, = \, g \, \la G_{RPA}^{(\alp)} \ra \,.
\ee
Finally, the quantity of interest is the dynamical structure
function $S^{(\alp)}(q,\omega)$ which is, 
at zero temperature, proportional to the imaginary part 
of the response function at positive energies:
\be 
 S^{(\alp)}(q,\omega)
   = -\frac{1}{\pi} \, \rm{Im} \, \chi_{RPA}^{(\alp)}(q,\omega)\,. 
\label{eq:sfunc}
\ee


\subsection{Response function for the spin-orbit case}

As an introduction to our full calculation we recall here some results
already obtained by Margueron \etal~\cite{Margueron06}.
When the spin-orbit force alone is included, the response function can then 
be written in the form (using $\hbar=c=1$):
\bwt
\bqr
\frac{\chihf}{\chia_{RPA}} & = &  1 - \wtua \chiz 
    + \wda \left[ \frac{q^2}{2} \chiz - 2 k_F^2 \chid \right]            \\
                           & + & 
  \left[ \wda k_F^2 \right]^2 \left[ \chidd - \cchi 
  + \left(\frac{m^* \omega}{k_F^2} \right)^2 \chizd
            -\frac{m^*}{6\pi ^2k_F}q^2 \chiz 
                              \right]   
  +  2 \left( \frac{m^* \omega}{q} \right)^2
          \frac{\wda}{1-\mkctpi \wda} \chiz\,. \nn
\label{eq:RPAresp}
\eqr
\ewt

In this expression $k_F$ is the Fermi momentum whereas $m^*$
denotes the effective mass of the nucleons.
The functions $\chiz$, $\chid$ and $\chiq$ are generalized free response
functions, defined in~\citerefdot{Garcia92} and written
in~\citeAppendix{app:sec:glf}, see eq.~(\ref{eq:chi2i}),
and $\chi_{HF}=g \,\chiz$.

The explicit expressions of the
$\wtua$ coefficients are given in~\citeAppendix{app:sec:wsocoef}
where the coupling between the $S=0$ and $S=1$ channels induced by the 
spin-orbit interaction can also be clearly seen.
It can be noted that the coefficients $\wtua$ are now complex
functions of $q$ and $\omega$ since different moments $\chi_{2i}$ enter
their expressions.
If we replace $\wtua$ in~\citeqdot{eq:RPAresp} by $\wua$ we obtain 
the results of~\citerefdot{Garcia92}, related to the central part
of the interaction, as it should be.


\subsection{Response function with the tensor part}

As already mentioned, the tensor interaction couples
vector channels with different spin projection while the spin-orbit
interaction couples the scalar ($S=0$) and vector ($S=1$) ones.
The consequence is that we obtain a non trivial system of coupled equations
for the RPA problem.
As an illustration let us consider explicitly the case of isospin $I=0$ and
$(S,M) = (1,1)$.
In that particular channel, the Bethe-Salpeter equation for
$G_{RPA}^{(1,1)}(k_1)$ (we omit the isospin index and
dependence on $q$ and $\omega$ for sake of simplicity)
exhibits terms which typically read:
\bi
   \item $G_{HF}(k_1) \la G_{RPA}^{(1,1)} \ra $\,,
   \item $G_{HF}(k_1) \la k^2 G_{RPA}^{(1,1)} \ra$\,,
   \item $k_1 Y^{(1)*}_{\mu} G_{HF}(k_1) \la
                k Y^{(1)}_{\mu} G_{RPA}^{(1,1)} \ra$\,,
   \item $k_1 Y^{(1)}_{\text{-}1} G_{HF}(k_1) \la G_{RPA}^{(1,0)} \ra$\,,
   \item $G_{HF}(k_1) \la k^2 Y^{(1)}_{\text{-}1}
             Y^{(1)}_{\text{-}1} G_{RPA}^{(1,\,\text{-}1)} \ra$\,.
\ei
Thus, the determination of $\la G_{RPA}^{(1,1)} \ra$ 
requires the knowledge of some other unknown quantities.
This leads to a large system of coupled equations for: 

\bi
  \item $\la                   G_{RPA}^{(1,M)} \ra$\,,
  \item $\la k^2               G_{RPA}^{(1,M)} \ra$\,, 
  \item $\la k Y^{(1)}_0       G_{RPA}^{(1,M)} \ra$\,,
  \item $\la k^2 |Y^{(1)}_0|^2 G_{RPA}^{(1,M)} \ra$\,, 
  \item $\la k^2 |Y^{(1)}_1|^2 G_{RPA}^{(1,M)} \ra$ for $M=-1,0,1$\,,
\ei
and
\bi
  \item $\la k^2 Y^{(1)}_1    Y^{(1)}_1    G_{RPA}^{(1,1)}  \ra$\,,
  \item $\la k^2 Y^{(1)}_{\text{-}1} Y^{(1)}_{\text{-}1}
                                           G_{RPA}^{(1,\,\text{-}1)} \ra$\,,
  \item $\la k   Y^{(1)}_1                 G_{RPA}^{(1,1)}  \ra$\,,
  \item $\la k   Y^{(1)}_{\text{-}1}       G_{RPA}^{(1,\,\text{-}1)} \ra$\,,
  \item $\la k   Y^{(1)}_1                 G_{RPA}^{(1,0)}  \ra$\,,
  \item $\la k   Y^{(1)}_{\text{-}1}       G_{RPA}^{(1,0)}  \ra$\,,
\ei
that is 21 unknown quantities. 
Fortunately, since the multipole expansion of $G_{HF}$ only implies
terms with $Y^{(L)}_0$, the integration over $k_1$ cancels all terms 
of the form $\la f(k)\prod_{\{M,M',...\}}Y^{(1)}_M Y^{(1)}_{M'}... G_{HF} \ra$ 
with $M+M'+ ... \neq 0$.
Moreover some unknown quantities can be expressed through
the others and we can reduce the 21 coupled equations
to three systems of four coupled equations for the following variables:
\begin{align}
&\la G_{RPA}^{(1,M)} \ra\,,~~~
\la k^2 G_{RPA}^{(1,M)} \ra\,,~~~
\la k Y^{(1)}_0 G_{RPA}^{(1,M)} \ra\,,~~~ \nonumber \\[1mm]
&\la k^2 |Y^{(1)}_0|^2 G_{RPA}^{(1,M)}
\ra - \la k^2 |Y^{(1)}_1|^2 G_{RPA}^{(1,M)} \ra\,, \nonumber
\end{align}
for each value of $M$.
The calculations are straightforward but tedious.
The same procedure has to be repeated for isospin $I=1$ for $S=1$.
Indeed, only the channels with $S=0$ are less involved.

The results are now quoted for each spin-isospin channel in a form that
exhibits the symmetry properties appearing in~\citeTable{tab:phme_tens}:
\bwt
\bi
   \item For  $S=0, I=0$:
\bqr
\frac{\chihf}{\chi^{(0,0)}_{RPA}} 
 & = & \, 1 \, - \, {\wt {\wt W}}^{(0,0)}_1 \chiz 
            \, + \, \wzzd \left( \frac{q^2}{2} \chiz - 2 k_F^2 \chid \right)
            \, + \, \left[ \wzzd k_F^2 \right]^2 \crochi                \nn \\
 & + & \, 2 \chiz \moqd \frac{\wzzd}{1-\mkctpi \wzzd}\,.
\eqr
   \item For  $S=0, I=1$:
\bqr
\frac{\chihf}{\chi^{(0,1)}_{RPA}} 
 & = & \, 1 \, - \, {\wt {\wt W}}^{(0,1)}_1 \chiz 
       \, + \, \wzud \left( \frac{q^2}{2} \chiz -  2 k_F^2 \chid \right)    
       \, + \, \left[ \wzud k_F^2 \right]^2 \crochi                     \nn \\
 & + & \, 2 \chiz \moqd \frac{\wzud}{1-\mkctpi \wzud}\,.
\eqr     
   \item For $S=1, I=0, M = \pm 1$:
\bqr
\frac{\chihf}{\chi^{(1, 0,\pm 1)}_{RPA}} 
 & = & \, \left[ 1 + \tmc \mkcdpi \right]^2 
       \, - \, {\wt {\wt W}}^{(1,0,\pm 1)}_1 \chiz                          \\
 & + & \, \left[ \wuzd - \tmc \right]  
       \left\{ \frac{q^2}{2} \chiz \left[1+\tmc\mkcpi\right] - 2 k_F ^2 \chid
       + \tmc \mkfpi \left( \chiz - \chid \right) \right\}              \nn \\
 & + & \, \left[ \wuzd - \tmc \right]^2 k_F^4 \crochi                   \nn \\
 & + & \, 2 \moqd \frac{\ttpXa}{1-\mkctpi \left[ \ttmXa \right]}\,\chiz \nn \\
 & + &  \left\{ 2 m^*\omega \Xa + \dmkctpi \left[ \Xa \right]^2
        \left[ \frac{q}{2} + \moq \right]^2 \right\}
        \frac{\chiz }{1-\mkctpi \left[ \ttmXa \right]}\,.       \nn
\eqr
   \item For $S=1, I=0, M = 0$:
\bqr
\frac{\chihf}{\chi^{(1, 0,0)}_{RPA}} 
 & = & \, \left[ 1 - \pdemi \tmc \mkcdpi \right]^2 
       \, - \, {\wt {\wt W}}^{(1,0,0)}_1 \chiz                           \\
 & + & \, \left[ \wuzd + \pdemi \tmc \right]
           \left\{ \frac{q^2}{2} \chiz \left[ 1 - \tmc \mkcdpi \right]
       -  2k_F ^2 \chid - \tmc \mkfdpi \left(\chiz-\chid\right) \right\} \nn\\
 & + & \, \left[ \wuzd + \pdemi \tmc \right]^2 k_F^4 \crochi             \nn\\
 & + & \, 2 \moqd \frac{\ttpXb}{1-\mkctpi [ \ttmXb ]} \, \chiz           \nn\\
 & + & \, \left\{ - 2 m^* \omega \Xb + \dmkctpi \left[ \Xb \right]^2
          \left[ \frac{q}{2}-\moq \right]^2 \right\}
                            \frac{\chiz}{1-\mkctpi [\ttmXb]}\,. \nn
\eqr
   \item For $S=1, I=1, M = \pm 1$:
\bqr
\frac{\chihf}{\chi^{(1, 1,\pm 1)}_{RPA}} 
 & = & \, \left[ 1 + 3 \tm \mkcdpi \right]^2 
       \, - \, {\wt {\wt W}}^{(1,1,\pm 1)}_1 \chiz                         \\
 & + & \, \left[ \wuud - 3 \tm \right]
       \left\{ \frac{q^2}{2} \chiz \left[ 1 + 3 \tm \mkcpi \right]
       -  2k_F ^2 \chid + 3\tm \mkfpi \left(\chiz-\chid\right) \right\} \nn \\
 & + & \, \left[ \wuud - 3 \tm \right]^2 k_F^4 \crochi                  \nn \\
 & + & \, 2 \moqd \frac{\ttpXc}{1-\mkctpi [ \ttmXc ]} \, \chiz          \nn \\
 & + & \, 2 \moqd \mkctpi \left[ \Xc \right]^2
              \frac{\chiz }{1-\mkctpi [ \ttmXc ]}\,.  \nn
\eqr
   \item For $S=1, I=1, M = 0$:
\bqr
\frac{\chihf}{\chi^{(1, 1,0)}_{RPA}} 
 & = & \, \left[ 1 - \ptdemi \tm \mkcdpi \right]^2
       \, - \, {\wt {\wt W}}^{(1,1,0)}_1 \chiz                      \\
 & + &  \left[ \wuud + \ptdemi \tm \right]
        \left\{ \frac{q^2}{2} \chiz \left[ 1 - 3 \tm \mkcdpi \right]
       -  2 k_F ^2 \chid - \ptdemi \tm \mkfpi
                          \left( \chiz - \chid \right) \right\}      \nn \\
 & + & \, \left[ \wuud + \ptdemi \tm  \right]^2 k_F^4 \crochi        \nn \\
 & + & \, 2 \moqd \frac{\ttpXd}{1 - \mkctpi [ \ttmXd ]} \chiz        \nn \\
 & + & \, 2 \moqd \mkctpi \left[ \Xd \right]^2
         \frac{\chiz}{1-\mkctpi [ \ttmXd ]}\,.   \nn
\eqr
\ei

The coefficients ${\wt {\wt W}}^{(S,I)}_1$ for $S=0$ or
${\wt {\wt W}}^{(S,I,M)}_1$ for $S=1$ are defined as
\bqr
{\wt {\wt W}}^{(0,0)}_1       \, & = & \, W^{(0,0)}_1 \, + \, 9 \wsod q^4
                    \frac{\bdbt}{1+q^2(\bdbt) (\wuzd  - (7 t_e+13t_o))}     \\
{\wt {\wt W}}^{(0,1)}_1       \, & = & \, W^{(0,1)}_1 \, + \,   \wsod q^4
                    \frac{\bdbt}{1+q^2 (\bdbt) [\wuud + 3 \tm]}             \\
{\wt {\wt W}}^{(1,0,\pm 1)}_1 \, & = & \, {\wt W}^{(1,0,\pm 1)}_1 + K_1 + K_2  
	     - \, t_o^2 \, \frac{24m^*k_F^5}{5\pi^2}                    \nn \\ 
	& + & \, 9 (t_e+3t_o)^2 q^4 (\beta_5 - 2 \beta_8 + \beta_7)
	       + 2 \Xa \left[ \frac{q}{2} + \moq \right]^2                  \\
{\wt {\wt W}}^{(1,0,0)}_1 \, &=& \, {\wt W}^{(1,0,0)}_1-\pdemi K_1+\pquar K_2
	      - \, t_o^2 \frac{48m^*k_F^5}{5\pi^2}
		        + 2 \Xb \left[ \frac{q}{2}- \moq \right]^2          \\
{\wt {\wt W}}^{(1,1,\pm 1)}_1 \, &=& \, {\wt W}^{(1,1,\pm 1)}_1+3K'_1+9K'_2 
	      - \, \tm^2 \frac{6m^*k_F^5}{5\pi^2}                       \nn \\
	& + & \, 9 \tm^2 q^4 (\beta_5 - 2 \beta_8 + \beta_7) + 2 \Xc \moqd  \\
{\wt {\wt W}}^{(1,1,0)}_1 \,
              & = & \, {\wt W}^{(1,1,0)}_1 - \ptdemi K'_1 + \tfrac{9}{4} K'_2
	      - \, \tm^2 \frac{12m^*k_F^5}{5\pi^2} + 2 \Xd \moqd\,.
\eqr
The functions $\beta_i$ are given in~\citeAppendix{app:sec:glf}
and the coefficients $K_i$, $K_i'$ and $X^{(S,I,M)}$ are
\bqr
K_1 & = & 2 \tpc q^2 + 6 \moqd \tmc\,,                                      \\
K_2 & = & - \tmc^2 \, \mkcpi \left[ \frac{9}{10} k_F ^2 + \frac{3}{8} q^2
               - \tdemi \moqd \right]\,,                                    \\
K'_1 & = & 2 \tp q^2 + 6 \moqd \tm\,,                                       \\
K'_2 & = & - \tm^2 \, \mkcpi \left[ \frac{9}{10} k_F ^2 + \frac{3}{8} q^2
               - \tdemi \moqd \right]\,,
\eqr
and
\bqr
X^{(1,0,\pm 1)} \, & = &
 \, 36 t_o^2 q^2 \frac{\bdbt}{1 + q^2(\bdbt) ( \wuzd+\pdemi(t_e-5t_o))}\,, \\
X^{(1,0,0)}     \, & = &
 \, 72 t_o^2 q^2 \frac{\bdbt}{1 + q^2(\bdbt) ( \wuzd + 5t_e+23t_o)}\,,    \\
X^{(1,1,\pm 1)} \, & = &
 \,  9 \tm^2 q^2 \frac{\bdbt}{1 + q^2(\bdbt) ( \wuud + \ptdemi \tm )}\,,  \\
X^{(1,1,0)}     \, & = &
 \, 18 \tm^2 q^2 \frac{\bdbt}{1 + q^2(\bdbt) ( \wuud -       9 \tm )}\,.
\eqr
\ewt

It is important to note that the above response functions have exactly
the same structure as in HF, with or without the spin-orbit interaction.
Moreover we can see that the response function depends on different
linear combinations of the parameters $t_e$ and $t_o$ that will lead
to non trivial effects.


\section{Dynamical structure function with tensor contribution}
\label{sect:results}

As an example for the effect of a zero-range tensor force in the p-h
interaction, we have calculated the nuclear responses in $(S,I,M)$
channels of symmetric infinite nuclear matter for parametrization T44
of the Skyrme interaction built by Lesinski \etal~\cite{Lesinski07}.
The parameters of this force are given in Table~\ref{tab:t44}.
The dynamical structure functions
$S^{(\alp)} (\romq,\omega)$ (defined in \citeqdot{eq:sfunc})
calculated for $q=k_F$
and at the saturation density $\rho = \rho_\mathrm{sat}
=0.16~\mathrm{fm}^{-3}$ are shown on \citeFigure{fig:T44}.
To clearly isolate the effect of the tensor part of the force,
the functions $S^{(\alp)} (\romq,\omega)$ are plotted for two cases:
a) with no tensor contribution but including the spin-orbit part;
b) with the full force.
The first case allows to compare with the previous results
of Margueron \etal~\cite{Margueron06} but are presented here for the
four channels of the symmetric nuclear matter.

\begin{figure}[htbp]
      {\includegraphics[width=0.95\columnwidth,clip]{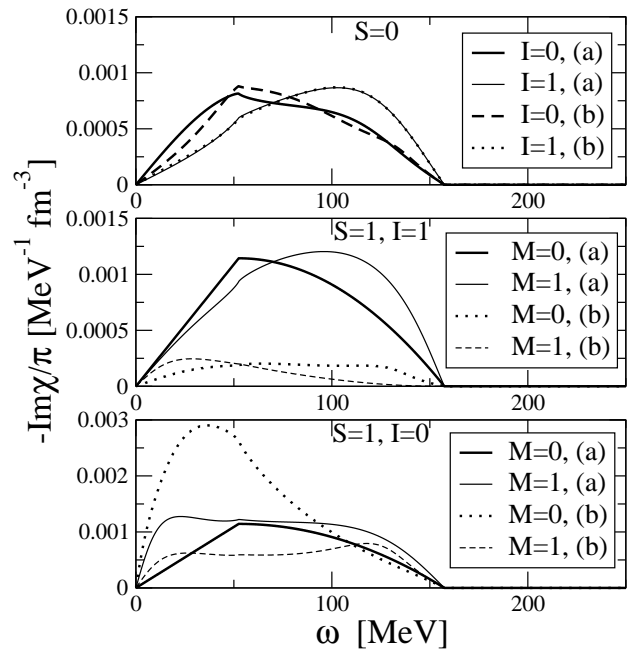}
       \caption{Dynamical structure functions $S^{(\alp)} (\romq,\omega)$
                calculated for the Skyrme tensor parametrization
                T44~\cite{Lesinski07}.
                The upper (middle and lower respectively) part of the Figure 
                concerns the channels $S=0$, $S=1$ and $I=0$, $S=1$ and $I=0$.
		 For each $(S,I,M)$ channel the responses
                (in MeV$^{-1}$\,fm$^{-3}$) 
                are plotted as a function of $\omega$ (in MeV).
                All the responses are calculated for $q=k_F$ at saturation
                density. Cases (a) and (b) are discussed in the text.}
       \label{fig:T44}}
\end{figure}

\begin{table}[htbp]
\begin{tabular}{rrrrrrr}
\hline
\hline
\mcc{$t_0$} & \mcc{$x_0$} & \mcc{$t_1$} & \mcc{$x_1$} & \mcc{$t_2$} &
\mcc{$x_2$} \\
\hline
\,-2485.67 & \,\,0.721557 & \,494.477 & \,-0.661848 & \,-337.961 & \,-0.803184
 \\
\hline
\mcc{$t_3$} & \mcc{$x_3$} & \mcc{$\gamma$} & \mcc{$W_0$} & \mcc{$t_e$} &
\mcc{$t_o$} \\
\hline
13794.7 & 1.175908 & \mcc{1/6} & 161.367 & 173.661 & 7.17383 \\
\hline
\hline
\end{tabular}
\caption{ \label{tab:t44} Parameters of the interaction T44~\cite{Lesinski07}
as defined in equations (\ref{eq:Skyrme:central}),
(\ref{eq:Skyrme:LS}) and (\ref{eq:Skyrme:tensor}).}
\end{table}

In $S=0$ channel, the tensor terms do not affect qualitatively the response.
Strictly speaking, this comment only applies to the case
of the T44 tensor parametrization but several tests performed using other
Tij tensor interactions discussed in~\citerefdot{Lesinski07} exhibit the same
qualitative behaviour.
The situation is quite different in $S=1$ channels: the effect from the tensor
terms is large whatever the value of the spin projection $M$ is.
Actually, depending on the values of the transferred momentum
$q$ and the density $\rho$,  the response functions increase significantly
and diverge at finite $q$ for a certain critical
density $\rho_c$. This divergence
reveals the presence of instabilities observed in nuclei \cite{Les06a},
with the appearance of domains with typical size of the order of $2\pi/q$.
Even if a one-to-one correspondence between infinite matter and finite
nuclei is obviously not correct, it remains that the center of a
nucleus explores, because of fluctuations, not only the saturation
density but also some larger values for which one may observe a divergence
of the response functions, and then, possibly, the appearance of
finite-size instabilities in the nucleus.


\section{Summary and conclusions}
\label{sect:conclusions}

We have derived the contribution from the tensor terms in the Skyrme
effective interaction to the RPA response
function. We have shown that the formal structure of the response
function is the same as without tensor terms, although with the
latter, all channels are coupled in a non trivial way. The simple example
presented here, using the interaction T44, shows that the effects
of the tensor contributions are strong in vector channels.

We have shown that the dynamical structure functions $S^{(\alpha)}(q,\omega)$
becomes large for finite values of $q$. This indicates the vicinity
of a pole related with a finite size instability for given values
of $q$ and $\rho$ in infinite matter and, possibly, in finite
nuclei. A systematic study of the critical
densities is in progress in order to determine if the link
between the divergences of $\chi_{RPA}$ and the instabilities
encountered in nuclei at the Hartree-Fock approximation is robust.

Another important point under study is the identification, directly from
the Skyrme energy functional, of the origin of each tensorial contribution
in the response functions.
In the same spirit, a detailed study of sum rules can enlighten the
contribution of the tensor
for various physical situations (see \cite{Lipparini}). 
Finally, applications to pure neutron matter can be very important (see by
instance~\cite{Marcos91,Fantoni01,Margueron02,Vidana02a,Vidana02b,
Margueron09,Isayev04,Beraudo04,Beraudo05,Rios05,Lopez-Val06,Krastev07,
Bordbar08}).
However, above formulae are no longer directly usable and have 
to be adapted to that specific case.
Work in that direction is also in progress.


\section*{Acknowledgments}

The authors thank J. Margueron and J. Navarro for fruitful discussions
about their previous work which constitutes an essential piece of the
present paper. 
We also thank Nguyen Van Giai for enlightening discussions,
and M.~Bender and T.~Duguet for a critical reading of the manuscript.
M.~M. acknowledges the hospitality of the Theory Group of the Institut
de Physique Nucl\'eaire de Lyon during the two postdoctoral years at the
University of Lyon.


\begin{appendix}


\section{Coupling constants of the Skyrme energy functional}
\label{app:sec:cpl}

The Skyrme energy density functional used in this article
is a functional of local densities
and currents (with $q=n,\,p$ for neutrons and protons),
{\em i.e.}
the particle densities $\rho_q (\br)$,
the kinetic densities $\tau_q (\br)$,
the current (vector) densities $\vec{j}_q (\br)$,
the spin (pseudovector) densities $\vec{s}_q (\br)$,
the spin kinetic (pseudovector) densities $\vec{T}_q (\br)$,
the spin-current (pseudotensor) densities  $J_{q,\mu \nu}(\br)$ and
the tensor-kinetic (pseudovector) densities $\vec{F}_{q} (\br)$
that have been defined in Lesinski \etal~\cite{Lesinski07},
where the definition of the vector spin current
density $\vec{J}^{(1)}(\br) \equiv \vec{J} (\br)$ was also recalled.
We will use also the isoscalar and isovector densities
defined from the proton and neutron densities respectively as:
$\rho_{0} (\br) = \rho_\mathrm{n} (\br) + \rho_\mathrm{p} (\br)$ and
$\rho_{1} (\br) = \rho_\mathrm{n} (\br) - \rho_\mathrm{p} (\br)$,
and similar for all other densities. 

The Skyrme energy functional representing central, tensor and spin-orbit
contributions is given by:
\bwt
\bqr
\label{eq:EF:full}
\mathcal{E}_{\text{Skyrme}}
& = &   \mathcal{E}_\mathrm{C}
      + \mathcal{E}_\mathrm{LS}
      + \mathcal{E}_\mathrm{T}
      \nn \\
& = & \int \! \mathrm d^3\mathbf r \sum_{t=0,1}
      \bigg\{ \,      C^\rho_t [\rho_0] \, \rho_t^2
              \, + \, C^{\Delta \rho}_t \, \rho_t \Delta \rho_t
              \, + \, C^\tau_t          \, ( \rho_t \tau_t - \vec{j}_t^2 )   
      \nn \\
&   &         \, + \, C^s_t [\rho_0] \, \vec{s}_t^2
              \, + \, C^{\nabla s}_t \, \left(\vnabla\cdot\vec{s}_t\right)^2 
              \, + \, C^{\Delta s}_t \, \vec{s}_t \cdot \Delta \vec{s}_t   \\
&   &         \, + \, C^{T}_t \Big(  \vec{s}_t \cdot \vec{T}_t
                 - \sum_{\mu, \nu = x}^{z} J_{t, \mu \nu} J_{t, \mu \nu} \Big)
              \, + \, C^{F}_t \Big[  \vec{s}_t \cdot \vec{F}_t
                  - \pdemi \Big( \sum_{\mu = x}^{z} J_{t,\mu \mu} \Big)^2
                  - \pdemi \sum_{\mu, \nu = x}^{z} J_{t, \mu \nu} J_{t,\nu\mu}
                \Big]
      \nn \\
&   &
              \, + \, C^{\nabla J}_t \, \big[ \rho_t \vnabla \cdot \vec{J}_t
                   + \vec{s}_t \cdot \big( \vnabla \times \vec{j}_t \big) \big]
      \bigg\} \,. \nn
\eqr
\ewt

All the coupling constants can be expressed as function of the parameters
of the Skyrme interaction given 
in~\citeqssdot{eq:Skyrme:central}{eq:Skyrme:LS}{eq:Skyrme:tensor}.
Some of the coupling constants are fully defined by the standard central part
of the Skyrme force:
$C^{\rho}_t        = A^{\rho}_t$\,, 
$C^{\Delta \rho}_t = A^{\Delta \rho}_t$\,,
$C^{\tau}_t        = A^{\tau}_t$\, and
$C^{s}_t           = A^{s}_t$\,,
or by its spin-orbit part:
$C^{\nabla J}_t = A^{\nabla J}_t$,
Two coupling constants only depend on the tensor part of the interaction:
$C^{\nabla s}_t = B^{\nabla s}_t$, and
$C^{F}_t        \, = \, B^{F}_t$.
Finally,
two coupling constants are the sum of contributions from both central and
tensor forces:
$C^{T}_t        = A^{T}_t        + B^{T}_t$ and
$C^{\Delta s}_t = A^{\Delta s}_t + B^{\Delta s}_t$.

The coupling constants of the Skyrme EDF which come from the central
and spin-orbit parts of the interaction are given in terms of the
parameters by:
\bqr
\label{eq:cpl:SF}
A_0^{\rho} \, &=& \,   \tfrac{3}{8}  t_0
                     + \tfrac{3}{48} t_3 \,\rhozg(\br)\,,               \nn \\
A_1^{\rho} \, &=& \, - \tfrac{1}{4}  t_0 \big( \pdemi + x_0 \big)
                     - \tfrac{1}{24} t_3 \big( \pdemi + x_3 \big)
                                              \, \rhozg (\br)\,,        \nn \\
A_0^{\Delta \rho} \, &=& \, - \tfrac{9}{64} t_1
                            + \tfrac{1}{16} t_2
                              \big( \tfrac{5}{4} + x_2 \big)\,,         \nn \\
A_1^{\Delta \rho} \, &=& \,   \tfrac{3}{32} t_1 \big( \pdemi + x_1 \big)
                        + \tfrac{1}{32} t_2 \big( \pdemi + x_2 \big)\,, \nn \\ 
A_0^{\tau} \, &=& \, \tfrac{3}{16} \, t_1
                   + \tfrac{1}{4} t_2\;\big(\tfrac{5}{4} + x_2 \big)\,, \nn \\
A_1^{\tau} \, &=& \, - \tfrac{1}{8} t_1 \big(\pdemi + x_1 \big)
                     + \tfrac{1}{8} t_2 \big(\pdemi + x_2 \big)\,,      \nn \\
A_0^{s}    \, &=& \, - \tfrac{1}{4}  t_0 \big( \pdemi - x_0 \big)
                     - \tfrac{1}{24} t_3 \big( \pdemi - x_3 \big)
                                         \, \rhozg (\br)\,,             \nn \\
A_1^{s}    \, &=& \, - \tfrac{1}{8}  t_0
                     - \tfrac{1}{48} t_3 \, \rhozg (\br)\,,             \nn \\
A_0^{\Delta s} \, &=& \, \tfrac{3}{32} t_1 \big( \pdemi - x_1 \big)
                       + \tfrac{1}{32} t_2 \big( \pdemi + x_2 \big)\,,  \nn \\
A_1^{\Delta s} \, &=& \, \tfrac{3}{64} t_1 + \tfrac{1}{64} t_2\,,       \nn \\
A_0^{T}    \, &=& \, - \tfrac{1}{8} t_1 \big( \pdemi - x_1 \big) \,
                     + \tfrac{1}{8} t_2 \big( \pdemi + x_2 \big)\,,     \nn \\
A_1^{T}    \, & = & \, - \tfrac{1}{16} t_1 + \tfrac{1}{16} t_2\,,       \nn \\
A_0^{\nabla J} \, &=& \, - \tfrac{3}{4}\, W_0\,, \nn \\
A_1^{\nabla J} \, &=& \, - \tfrac{1}{4}\, W_0\,. \nn
\eqr

Due to the density dependence of some coupling constants
it is also useful to define the following coefficients
which occur in the definition of the $W^{(\alp)}$ 
coefficients~\cite{RMP}:
\bqr
   A^{\rho}_t \, & = & \, A_t^{\rho,0} + A_t^{\rho,\gamma} \rhozg\,,  \nn \\
   A^{s}_t    \, & = & \, A_t^{s,0}    + A_t^{s,\gamma}    \rhozg\,.  \nn
\eqr
In each case where we have considered a single density dependent term, the
generalization to more than one density dependent term is straightforward
just adding new density dependent terms in the corresponding coupling
constants.

Finally, the coupling constants of the Skyrme EDF which come from the
tensor part of the interaction are given by
(Table I in~\cite{Per04a}):
\begin{alignat}{4}
\label{eq:cpl:tensor}
B^{T}_0 & = - \tfrac{1}{8} (t_e + 3 t_o)\,,                   & \qquad
B^{T}_1 & = \phantom{-} \tfrac{1}{8} (t_e - t_o)\,,                \nn \\
B^{F}_0 & = \phantom{-} \tfrac{3}{8} (t_e + 3 t_o)\,,         & \qquad
B^{F}_1 & = - \tfrac{3}{8} (t_e - t_o)\,,                          \nn \\
B^{\Delta s}_0 & = \phantom{-} \tfrac{3}{32} (t_e - t_o)\,,   & \qquad
B^{\Delta s}_1 & = - \tfrac{1}{32} (3 t_e + t_o)\,,                \nn \\
B^{\nabla s}_0 & =   \phantom{-} \tfrac{9}{32} (t_e - t_o)\,, & \qquad
B^{\nabla s}_1 & =  - \tfrac{3}{32} (3 t_e + t_o)\,.               \nn
\end{alignat}

%
%
\section{$W^{\alp}$ coefficients: central part of the force.}
\label{app:sec:wcoef}

With the coupling constants defined in the Skyrme energy density functional
and recalled in~\citeAppendix{app:sec:cpl}, the $W^{(\alp)}$ coefficients take 
the following expressions for symmetric nuclear matter 
($\rho_n = \rho_p = \pdemi \rho_0$ and $\rho_1=0$):
\bqr
\pquar W^{(0,0)}_1 & = &  2 A_0^{\rho,0} 
                          + A_0^{\rho,\gamma} \cogam \rhozg           \nn \\
                   & - &  \left[ 2 A_0^{\Delta \rho} 
                          + \pdemi A_0^{\tau} \right] q^2\,,         \nn \\
\pquar W^{(0,1)}_1 & = &  2 A_1^{\rho,0} + 2 A_1^{\rho,\gamma} \rhozg  \nn \\
                   & - &  \left[ 2 A_1^{\Delta \rho} 
                          + \pdemi A_1^{\tau} \right] q^2\,,         \nn \\
\pquar W^{(1,0)}_1 & = &  2 A_0^{s,0} + A_0^{s,\gamma} \rhozg          \nn \\
                   & - &  \left[ 2 A_0^{\Delta s} 
                          + \pdemi A_0^{T} \right] q^2\,,            \nn \\
\pquar W^{(1,1)}_1 & = &  2 A_1^{s,0} + 2 A_1^{s,\gamma} \rhozg        \nn \\
                   & - &  \left[ 2 A_1^{\Delta s} 
                          + \pdemi A_1^{T} \right] q^2\,,            \nn \\
\pquar W^{(0,0)}_2 & = &    A_0^{\tau}\,,                            \nn \\
\pquar W^{(0,1)}_2 & = &    A_1^{\tau}\,,                            \nn \\
\pquar W^{(1,0)}_2 & = &    A_0^{T}\,,                               \nn \\
\pquar W^{(1,1)}_2 & = &    A_1^{T}\,.                               \nn
\eqr

%
%
\section{${\wt W^{(\alp)}}$ coefficients: spin-orbit part of the force.}
\label{app:sec:wsocoef}

When the spin-orbit force is taken into account in the RPA formalism, 
one has to define with the help of the $\beta_i$ moments~\cite{Garcia92} 
given in the~\citeAppendix{app:sec:glf}:
\bi
   \item for $S=0$, the coefficients ${\wt W^{(S,I)}_1}$:
\bqr
{\wt W^{(0,0)}_1} & = &
   \wzzu + 9 \wsod q^4 \frac{\bdbt}{1+\wuzd q^2 (\bdbt)}\,,  \nn \\
{\wt W^{(0,1)}_1} & = &
 \wzuu +   \wsod q^4 \frac{\bdbt}{1+\wuud q^2 (\bdbt)}\,.  \nn
\label{eq:wtSI}
\eqr
   \item for $S=1$, the coefficients ${\wt W^{(S,I,M)}_1}$:
\bqr
{\wt W^{(1,0,1)}_1}  & = &
   \wuzu + \frac{9}{2} \wsod q^4 \frac{\bdbt}{1+\wzzd q^2 (\bdbt)}\,, \nn \\
{\wt W^{(1,0,0)}_1}  & = & \wuzu\,,                                   \nn \\
{\wt W^{(1,0,-1)}_1} & = &
 \wuzu + \frac{9}{2} \wsod q^4 \frac{\bdbt}{1+\wzzd q^2 (\bdbt)}\,,   \nn \\
{\wt W^{(1,1,1)}_1}  & = &
 \wuuu + \demi \wsod q^4 \frac{\bdbt}{1+\wzud q^2 (\bdbt)}\,,         \nn \\
{\wt W^{(1,1,0)}_1}  & = & \wuuu\,,                                   \nn \\
{\wt W^{(1,1,-1)}_1} & = &
 \wuuu + \demi \wsod q^4 \frac{\bdbt}{1+\wzud q^2 (\bdbt)}\,.         \nn
\label{eq:wtSIM}
\eqr
\ei

%
%
\section{Generalized Lindhard functions}
\label{app:sec:glf}

Following~\citerefdot{Garcia92}, the definition of the generalized free 
response functions is:
\be
\chi_{2i} (q) = \!\!\int\! \frac{\mathrm d^4 q_3}{2(2\pi)^4}\!
 \left[ \left(\frac{q_3^2}{k_F^2}\right)^i 
\!\!+ \left(\frac{(\bqt + \bq)^2}{k_F^2}\right)^i\right] G_{HF} ( q_3 , q )\,.
\label{eq:chi2i}
\ee

The explicit expression can be found in~\citerefdot{Garcia92} for $i=0,1,2$.
With $\chiz, \chid$ and $\chiq$ we can compute the different moments of
$G_{HF}$ occurring in the calculation. 
As Garcia-Recio \etal~\cite{Garcia92},
we introduce the functions $\beta_i$ as:
\bwt
\be
\left[ \,\beta_i, ~ i=0,\,8 \,\right] =
\!\int\! \frac{\mathrm d^4 \bq_3}{(2\pi)^4} \, G_{HF} 
     \left[ ~ 1\,,
            \,~ \frac{ \bq \cdot\bq_3}{q^2}\,,
            \,~ \frac{ \bq_3^2}{q^2}\,,
            \,~ \frac{(\bq \cdot \bq_3)^2}{q^4}\,,
            \,~ \frac{(\bq \cdot \bq_3)\bq_3^2}{q^4}\,,
            \,~ \frac{ \bq_3^4}{q^4}\,,
            \,~ \frac{(\bq \cdot \bq_3)^3}{q^6}\,,
            \,~ \frac{(\bq \cdot \bq_3)^4}{q^8}\,,
            \,~ \frac{(\bq \cdot \bq_3)^2 \bq_3^2}{q^6}
              ~ \right]\,,
\nn
\ee
\ewt
with the functions $\beta_i$ written as
\bqr
       \beta_0 & = & \chiz\,,                                   \nn \\[0.7mm]
 2 k   \beta_1 & = & (\nu - k) \chiz\,,                         \nn \\[0.7mm]
 4 k^2 \beta_2 & = & \chid - 2 k \nu \chiz\,,                   \nn \\[0.7mm]
 4 k^2 \beta_3 & = & (\nu - k)^2 \chiz - \mkspi\,,              \nn \\[0.7mm]
 8 k^3 \beta_4 & = & 2 k \nu (k -\nu) \chiz
                                    + (\nu - k)\chid+\mktpi k\,,\nn \\[0.7mm]
16 k^4 \beta_5 & = & \chiq - 4 k \nu \chid\,,                   \nn \\[0.7mm]
 8 k^3 \beta_6 & = & (\nu - k)^3 \chiz + (3 k - \nu) \mkspi\,,  \nn \\[0.7mm]
16 k^4 \beta_7 & = & (\nu - k)^4 \chiz                          \nn \\
               & - & \mkdpi \,\left[ k^2+\frac{1}{5}
                             +\frac{1}{3} (2k-\nu)^2 \right]\,, \nn \\[0.7mm]
16 k^4 \beta_8 & = & (\nu - k)^2 \chid
                               - 2 k \nu (\nu - k)^2 \chiz      \nn \\
               & - & \mkspi \, \left[ 1 + 2k (3k-\nu) \right]\,.\nn
\eqr
where $k=\tfrac{q}{2k_F}$ and $\nu=\tfrac{m^*\omega}{qk_F}$.

For completeness, we now quote the different moments of $G_{HF}$
encountered in the RPA-equations:
\begin{align}
\la G_{HF} \ra  &=  \beta_0\,,                            \nn \\
\la k\, Y_0^{(1)} G_{HF} \ra    &=  q   \sqrt{\tqpi} \beta_1\,, \nn \\
\la k^2 G_{HF} \ra &= q^2 \beta_2\,,                      \nn \\
\la k^2 \left|Y_0^{(1)}\right|^2 G_{HF} \ra  &= q^2 \tqpi \beta_3\,,     \nn \\
\la k^2 \left|Y_1^{(1)}\right|^2 G_{HF} \ra   &= q^2 \thpi (\bdbt)\,,    \nn \\
\la k^3 Y_0^{(1)} G_{HF} \ra &= q^3 \sqrt{\tqpi} \beta_4\,,   \nn \\
\la k^3 \left|Y_0^{(1)}\right|^2 Y_0^{(1)} G_{HF} \ra
       & = q^3 \left( \tqpi \right)^{3/2} \beta_6\,,      \nn \\
\la k^3 \left|Y_1^{(1)}\right|^2 Y_0^{(1)} G_{HF} \ra
       & = q^3 \sqrt{\tqpi} \thpi (\beta_4 - \beta_6)\,,  \nn \\
\la k^4\,G_{HF} \ra  & = q^4 \beta_5\,,                    \nn \\
\la k^4\left|Y_0^{(1)}\right|^4G_{HF}\ra &
                = q^4 \left(\tqpi\right)^2 \beta_7\,, \nn \\
\la k^4 \left|Y_1^{(1)}\right|^2 \left|Y_0^{(1)}\right|^2 G_{HF} \ra
       & = q^4 \frac{9}{32\pi^2} ( \beta_8 - \beta_7)\,,  \nn \\
\la k^4 \left|Y_1^{(1)}\right|^4 G_{HF} \ra 
       & = q^4 \frac{9}{64\pi^2} (\beta_5 - 2\beta_8 + \beta_7)\,. \nn
\end{align}


\end{appendix}


\bibliography{linear_response}


\newpage
\begin{sidewaystable}[p]
\caption{Particle-hole matrix elements for the tensor part of the effective force.
Only the vector channels are concerned in this case and a factor 
$2\delta_{S1} \delta_{S'1} \delta_{II'}$ is implicit for each term.}
\label{tab:phme_tens}
\begin{ruledtabular}
\begin{tabular}{ccccc} 
                     &&&&                                                                \\
\multicolumn{2}{c}{} & $M'=1$ & $M'=0$ & $M'=-1$                                         \\ 
                     &        &        &                                                 \\ \hline
                     &        &        &                                                 \\
\multirow{6}{12mm}{$M=1$} & $I=0$ & $  (t_e+5t_o) q^2$ & $3 q t_o \kum$ & $-3 (t_e+3t_o) \kum \kum$    \\ 
                          &       & $+ (t_e-5t_o) \brk$ &    $-6t_o\kum\kuz$ &                         \\
                          &       &&&                                                    \\ 
                          & $I=1$ & $3 \tp q^2$ & $3 \tm \kum \kuz$ & $3 \tm \kum \kum$  \\
                          &       & $+ 3 \tm \left[ \kuz \kuz + \kum \kuu \right]$ &&    \\
                          &       &&&                                                    \\ \hline
                          &       &&&                                                    \\ 
\multirow{6}{12mm}{$M=0$} & $I=0$ & $3 q t_o \kuu$ & $-\demi (t_e+5t_o) q^2$  & $3 t_o q \kum$          \\  
                          &       &$+6t_o\kuu\kuz$& $- \demi (t_e-5t_o) \brk$ &   $+6t_o\kum\kuz$       \\ 
                          &       &&&                                                    \\        
                          & $I=1$ & $-3 \tm \kuu \kuz$ & $-\tdemi \tp q^2$ & $-3 \tm \kum \kuz$ \\ 
                          &       && $- \tdemi \tm \brk$ &                                      \\ 
                          &       &&&                                                           \\ \hline
                          &       &&&                                                           \\        
\multirow{6}{12mm}{$M=-1$}& $I=0$ & $-3 (t_e+3t_o) \kuu \kuu$ & $3 q t_o \kuu$ & $(t_e+5t_o) q^2$      \\
                          &       &&$-6t_o\kuu\kuz$& $+ (t_e-5t_o) \brk$                               \\ 
                          &       &&&                                                           \\
                          & $I=1$ & $3 (t_e-t_o) \kuu \kuu$ & $3 \tm \kuu \kuz$ & $3 \tp q^2$          \\
                          &       &&& $+ 3 \tm \brk$                                            \\
                          &       &&&                                                           \\
\end{tabular}
\end{ruledtabular}
\end{sidewaystable}


\end{document}